\documentclass[10pt]{article}
\usepackage[utf8]{inputenc}
\usepackage{graphicx}
\usepackage[utf8]{inputenc}
\usepackage{amsmath}
\usepackage{amsfonts}
\usepackage{amssymb}
\usepackage{graphicx}
\usepackage{setspace}
\usepackage{authblk}
\usepackage{caption}
\usepackage{float}
\usepackage{cite}
\usepackage[left=2.0cm,right=2.0cm,top=2cm,bottom=2cm]{geometry}
\usepackage{color}
\usepackage{color,soul}
\usepackage{multicol}
\title{\textbf{Study of Einstein Equivalence Principle with Sagnac Effect in Lorentz and Galilean Frame}}

\author[1]{Shouvik Sadhukhan}

\affil[1]{\small{Department of Physics, Indian Institute of Space Science and Technology(IIST), P.O: Valiamala, Trivandrum - 695547, State: Kerala; India}}
\affil[1]{\small{Email: shouvikphysics1996@gmail.com}}

\date{Dated : \today}
\begin{document}
\maketitle

\begin{abstract}
Sagnac effect has been studied in terms of Gyroscopic system in both Lorentz frame as well as flat Einstein frame. The Einstein equivalence principle has been used to determine the phase shift due to pseudo force in the transformation from rotating earth frame to stationary frame. A polychromatic broadband source has been considered for the discussion. The Langevin-Landau-Lifschitz metric has been used during the incorporation of General theory of relativity. The square type sagnac interferometer has been used in theory establishment.\\

\end{abstract}

\begin{multicols}{2}
{

\section{Introduction}
General relativity and its explanations are known and used widely in many papers \cite{1}. Einstein's Equivalence principle is one of the general relativistic approach available in literature \cite{2,3}. It states the equivalence between the gravitational force and inertial mass \cite{4}. In other words, it can be tell that the equivalence principle basically provides a correspondence relation between gravitational force experienced by a massive system and pseudo force experienced by a body under non inertial frame of references \cite{5}. This equivalence has been reestablished with Sagnac effect and Coriolis force with gyroscopic interferometer \cite{6}.\par
The flat space time geometry has been used here which is also similar as Minkowskian geometry \cite{7}. In our work we have considered rotating frame of reference to provide pseudo force and hence the metric or line element gets a transformation due to rotation. This modified form of rotating geometry is named as Langevin-Landau-Lifschitz metric geometry \cite{8}. This metric space system can provide rotational pseudo force on a body under motion on this frame. Our present work in mainly focused on this metric space system \cite{9}.\par
During the development of Einstein gravity it was necessary to provide some link between Einstein curvature based gravity model and Newtonian gravitational force theory. Thus, the idea of linearized gravity has been brought in literature. The gravity is nothing but the effect of scale time curvature produced by mass and hence this force is force is equivalent to the pseudo effect of inertial mass \cite{10}. now large distance weak gravity can be compared with Newtonian gravity theory and thus the weak gravity approximation has been designed in theory. The space time curvature is nothing but a weak gravity perturbation happened due to mass of the system which also provides the gravitational wave as a solution of propagating wave. This linearized gravity can be equivalent to Sagnac effect under non inertial frame of reference \cite{11}.\par
Sagnac interferometer experiments are used widely in many measurements \cite{12,13,14,15}. Sagnac Interferometer is a square structured interferometric set up where three reflecting mirror is used (figure 2) \cite{16,17,18,19,20}. The position of source and detector remains perpendicular to each other. The whole system base can be provide gyroscopic motion \cite{21,22,23,24,25}. The rotation may be relativistic or non relativistic. In our present work we have also tried to incorporate polychromatic source spectral interferometry based of gyroscopic Sagnac Interferometer \cite{26,27}. The polychromatic source and its spectral content interference output can provide highly sensitive displacement measurements and hence this has been used to analyze the pseudo effect of rotation in our work \cite{28}. Some details analysis of Einstein gravity can also be found in the references \cite{29,30,31,32,33,34,35,36,37,38,39}.\par
The paper has been proceeded accordingly, Section 2 we have given the mathematical overviews and theoretical analysis of Sagnac effect using Lorentz and Geometric frame. In section 3 we have mage our conclusion of our work.

\section{Mathematical analysis of Gravitational Equivalence of Sagnac Effect}
In this section we discuss the mathematical analysis of special theory of relativity with white light interferometry. We start from the standard theory for relativistic formalism with temporal delay in dynamic interferometer. The Minkowski space time or in Lorentz frame the line element using cylindrical coordinate system can be written as follows. \cite{1,2,3,4,5,6,7,8,9}
\begin{equation}
    ds^2=c^2dt^2-dr^2-r^2d\phi^2-dz^2
\end{equation}
Our experimental study contains a planner gyroscope and hence with respect to a rotating disc we should have symmetry around z axis. Thus, the line element should be reduced to the following equation.
\begin{equation}
    ds^2=c^2dt_{0}^2-R^2d\phi^2
\end{equation}
Here, $t_{0}$ represent the time with respect to inertial frame and $R$ is the measure of radius of the gyroscope. Now after applying rotation in the frame, we must have $\phi=\phi_0-\omega t$ where $\omega$ is the angular speed of the gyroscope. Hence, substituting this into the reduced line element, we can find the following form.
\begin{equation}
    ds^2=(c^2-\omega^2R^2)dt^2-R^2d\phi^2-2R^2\omega d\phi dt
\end{equation}
Now comparing this relation with the generalized line element $ds^2=g_{00}dt^2-g_{\phi\phi}d\phi^2-2g_{0\phi}\omega d\phi dt$ we can write the differential time for light as follows. For light we must have to consider null geometry and hence we have $ds=0$.
\begin{equation}
    dt=\frac{-g_{0\phi}d\phi\pm\sqrt{(g_{0\phi}^2d\phi^2-g_{\phi\phi}g_{00}d\phi^2)}}{g_{00}}
\end{equation}
For light like solutions with future directed system we must have $dt>0$. For clock wise or anti-clock wise rotation we must have $d\phi>0$ and $d\phi>0$ respectively. Hence the temporal difference due to rotation can be established as follows.
\begin{equation}
    \Delta t=-2\oint_{l}{\frac{g_{0\phi}}{g_{00}}Rd\phi}
\end{equation}
Here $l$ is the circumference of the gyroscope. For proper time frame the relation should be $d\tau=\sqrt{g_{00}}dt$. Hence we can write as follows.
\begin{equation}
    \Delta\tau=-2\sqrt{g_{00}}\oint_{l}{\frac{g_{0\phi}}{g_{00}}Rd\phi}
\end{equation}
Now converting these generalized form into real line elements, we can re-write them as follows.
\begin{equation}
    \Delta t=4\frac{\pi\omega R^2}{c^2(1-\frac{R^2\omega^2}{c^2})}
\end{equation}
and,
\begin{equation}
    \Delta \tau=4\frac{\pi\omega R^2}{c^2\sqrt{(1-\frac{R^2\omega^2}{c^2})}}
\end{equation}
The pictorial diagram for the relativistic frame can be given as follows. 
\begin{figure}[H]
\centering
\begin{minipage}[b]{0.5\textwidth}
    \includegraphics[width=\textwidth]{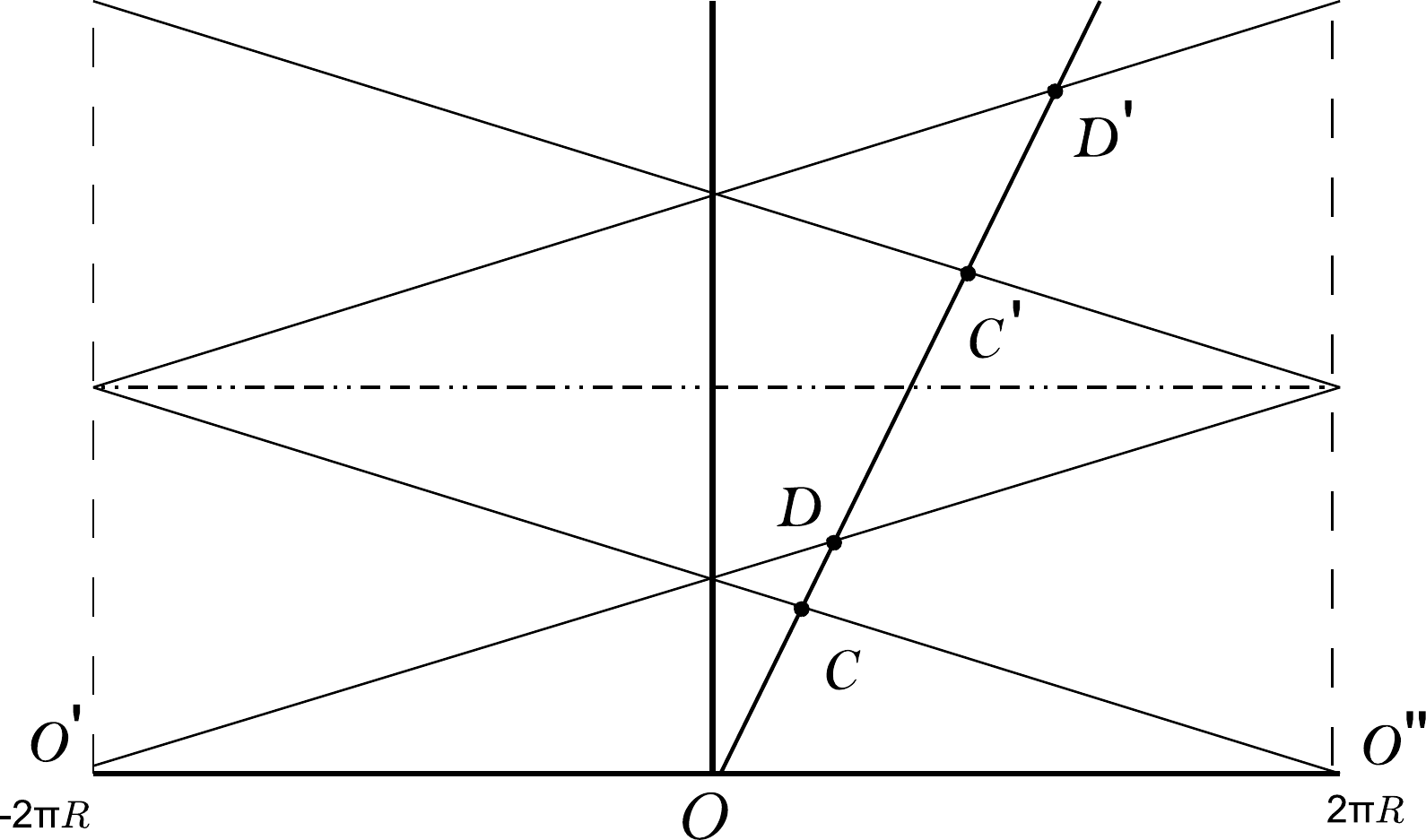}
    \caption{Rotating Relativistic Frame. Here, C and D are before rotation and C' and D' are after rotation. O is the present frame, O' is past and O'' is future frame reference point.}
\end{minipage}
\end{figure}
We are now providing our experimental diagram for the polychromatic gyroscopic Sagnac Interferometer. Here, we have used the Lorentz frame and hence the input signal should stay in broad band radio frequency region. The figure is as follows. \cite{13}
\begin{figure}[H]
\centering
\begin{minipage}[b]{0.5\textwidth}
    \includegraphics[width=\textwidth]{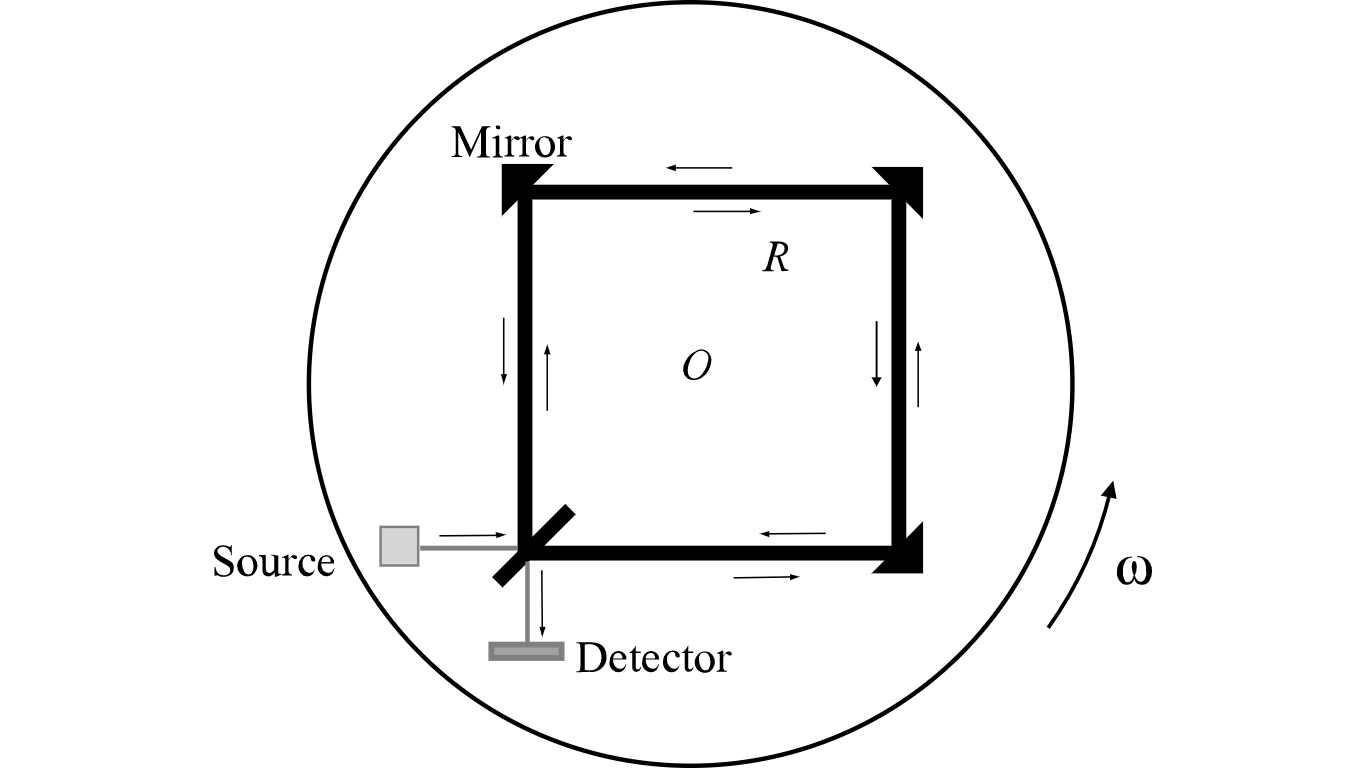}
    \caption{Sagnac Interferometer setup with the Gyroscopic table}
\end{minipage}
\end{figure}
In the experiments of relativistic sagnac effect we have used the broadband polychromatic source of radio frequency ranges. The central wavelength of that wave has been considered 6.12 m and bandwidth 2.38 m. The angular motions have been taken of the range $10^7$ rps and the radius is in meter range. Now we are providing the mathematical basis of our work that can produce the results discussed in this paper. Primarily in this work we have discussed the system with Gaussian optical beam source. In general we have considered the Gaussian beam with central peak intensity $H_0$ and the spectral domain intensity has been used as equation 1. (Spectral domain intensity should be represented ) Here we shall provide different angular motion in the table to calibrate the spectral switch with motion. \cite{1,2,3} Suppose the input spectral domain field is $\varepsilon_0(\lambda)$. Therefore, the input intensity given in equation \ref{1} can be rewritten as follows. \cite{27,28}
\begin{equation}\label{3}
    I_0(\lambda)=\varepsilon^{*}_0(\lambda)\varepsilon_0(\lambda)
\end{equation}
Now we have considered that our input spectral is Gaussian in nature and that's why the intensity is a Gaussian spectral content input. We have assumed that the spectral content input may have asymmetry as well as excess in nature. Thus, the input signal can be written as follows.
\begin{equation}\label{1}
    I_0 (\lambda)=H_0 \exp{(-\frac{1}{2}(\frac{\lambda-\lambda_0}{\Delta\lambda})^2)}g(\lambda)
\end{equation}
The term $g(\lambda)$ brings the asymmetry in the wave form. we have taken that function as in following equation.
\begin{equation}\label{2}
\begin{split}
    g(\lambda)=1+\frac{A}{6}((\frac{\lambda-\lambda_0}{\Delta\lambda})^2-3(\frac{\lambda-\lambda_0}{\Delta\lambda}))\\-\frac{C}{24}((\frac{\lambda-\lambda_0}{\Delta\lambda})^4-6(\frac{\lambda-\lambda_0}{\Delta\lambda})^2+3)
    \end{split}
\end{equation}
The constant $A$ is the asymmetry factor and $C$ is the excess. The wave form should follow the following condition.
\begin{center}
    $g(\lambda) \rightarrow
    \begin{cases}
      =1 & \text{for symmetric wave form spectrum} (A=B=0)\\
      \neq 1 & \text{for asymmetric wave form spectrum} (A\neq B\neq 0)
    \end{cases}$  
\end{center}
For our present work we have used both the symmetric and non-excess singlet Gaussian Beam in spectral domain and hence for the first part of our work we must have $A = B = 0$ and second part we have varied $A$ and $B$ with $A\neq B\neq 0$. We have used the frequency domain of the spectrum in this present work. The amplitude on the two arms of the interferometer can be taken as $a(\nu)$ and $b(\nu)$ and both of them are dependent on the transmittance and reflectance of the Beam splitter. We have considered the complex degree of coherence for this system is 1. therefore, the wave field for the two arms of the Interferometer can be taken as equation 3 and 4. ($a(\nu)=rt\exp{(ikL_1)}$ and $b(\nu)=tr\exp{(ikL_2)}$)
\begin{equation}\label{4}
   \varepsilon _1(\lambda)=a(\lambda)\varepsilon _{0}(\lambda)
\end{equation}
And,
\begin{equation}\label{5}
   \varepsilon _2(\lambda)=b(\lambda)\varepsilon _{0}(\lambda)
\end{equation}
The Field after interference will be as in next equation.
\begin{equation}\label{6}
    \varepsilon (\lambda)=\varepsilon_1 (\lambda)+\varepsilon_2 (\lambda)
\end{equation}
The intensity of the output can be found with $I(\lambda)=\varepsilon^{*} (\lambda) \varepsilon (\lambda)$. Finally the spectral domain intensity output with arbitrary phase $\theta$, can be written as follows.\cite{1,2,3,23}
\begin{equation}\label{7}
     I(\lambda)=H_0 \exp{(-\frac{1}{2}(\frac{\lambda-\lambda_0}{\Delta\lambda})^2)}(1+\cos{(\theta)})
\end{equation}
Therefore, all those equation will be used to determine the interfered intensity spectrum. This kind of Interferometer has same arm lengths for both directions of the interfering lights. The phase difference comes only with the rotation of the set up bench. Therefore in every cases we have considered $L_1=L_2$. Now we present calculation we shall only provide the results for Gaussian beam. We can not introduce the geometrical path difference in Sagnac Interferometer due to its structure. The rotation is the only process that can provide optical path difference and thus we can get spectral shift due to Gyroscopic motion of the interferometric set up. The mother equation for the whole work is the equation \ref{7}. Consider the relativistic angular velocity is $\omega$ and the tangential velocity due to this rotation is $\omega R$. Here $R$ is the radius of the interferometric set up.\par From the reference \cite{23} we can write the phase difference due to optical path difference with the motion as follows using the relativistic temporal delay derived above. The local time delay provides the following phase difference.
\begin{equation}\label{8}
    \theta=\frac{8\pi^2}{c}\frac{R^2\omega}{\lambda(1-\frac{\omega^2 R^2}{c^2})}
\end{equation}
The proper time delay provides the path difference as follows.
\begin{equation}\label{9}
    \theta=\frac{8\pi^2}{c}\frac{R^2\omega}{\lambda\sqrt{(1-\frac{\omega^2 R^2}{c^2})}}
\end{equation}
The path and phase difference we found are due to the rotation of gyroscope. Hence, there must have some kind of Doppler shift in the polychromatic wave lenghts. For the tangential velocity $\omega R$ we have the Doppler shift as follows.
\begin{equation}
    z=\gamma (1+\frac{\omega R}{c})-1=\frac{\Delta\lambda}{\lambda}
\end{equation}
For any wavelength $\lambda_i$, the Doppler shifted wavelength $\lambda_f$ can be written as follows.
\begin{equation}
    \lambda_f=\lambda_i\sqrt{(\frac{1+\frac{\omega R}{c}}{1-\frac{\omega R}{c}})}=\lambda_i\gamma (1+\frac{\omega R}{c})
\end{equation}
Here all $\lambda'$s can be substituted with $\lambda_i$ and we can re-write the input signal in relativistic frame as follows.
\begin{equation}
    I_0 (\lambda)=H_0 \exp{(-\frac{1}{2}(\frac{\lambda-\lambda_0}{\Delta\lambda})^2)(\frac{1+\frac{\omega R}{c}}{1-\frac{\omega R}{c}})}G(\lambda)
\end{equation}
Here $G(\lambda)$ is the relativistic form of $g(\lambda)$ which can be written as follows.
\begin{equation}
\begin{split}
    G(\lambda)=1\\+ \frac{A}{6}((\frac{\lambda-\lambda_0}{\Delta\lambda})^2(\frac{1+\frac{\omega R}{c}}{1-\frac{\omega R}{c}})-3(\frac{\lambda-\lambda_0}{\Delta\lambda})\sqrt{(\frac{1+\frac{\omega R}{c}}{1-\frac{\omega R}{c}})})\\-\frac{C}{24}((\frac{\lambda-\lambda_0}{\Delta\lambda})^4(\frac{1+\frac{\omega R}{c}}{1-\frac{\omega R}{c}})^2-6(\frac{\lambda-\lambda_0}{\Delta\lambda})^2(\frac{1+\frac{\omega R}{c}}{1-\frac{\omega R}{c}})+3)
    \end{split}
\end{equation}
Where the conditions for $G(\lambda)$ can also be written like this.
\begin{center}
    $G(\lambda) \rightarrow
    \begin{cases}
      =1 & \text{for symmetric wave form spectrum} (A=B=0)\\
      \neq 1 & \text{for asymmetric wave form spectrum} (A\neq B\neq 0)
    \end{cases}$  
\end{center}
Now the interference intensity follows the above equations. This can be written as follows. We have shown the intensity in terms of both local and proper time frame. For local time frame the intensity is as follows.
\begin{equation}
\begin{split}
    I_0 (\lambda)=H_0 \exp{(-\frac{1}{2}(\frac{\lambda-\lambda_0}{\Delta\lambda})^2)(\frac{1+\frac{\omega R}{c}}{1-\frac{\omega R}{c}})}G(\lambda)\\(1+\cos{(\frac{8\pi^2}{c}\frac{R^2\omega}{\lambda(1-\frac{\omega^2 R^2}{c^2})})})
\end{split}
\end{equation}
For proper time frame we can write the intensity as follows.
\begin{equation}
\begin{split}
    I_0 (\lambda)=H_0 \exp{(-\frac{1}{2}(\frac{\lambda-\lambda_0}{\Delta\lambda})^2)(\frac{1+\frac{\omega R}{c}}{1-\frac{\omega R}{c}})}G(\lambda)\\(1+\cos{\frac{8\pi^2}{c}\frac{R^2\omega}{\lambda\sqrt{(1-\frac{\omega^2 R^2}{c^2})}}})
\end{split}
\end{equation}
Now using this spectral content intensity we can calibrate the Doppler shift also with spectral switch shift. The details of the analysis have been given in physical analysis section. In the discussed set up we don't need to incorporate Coriolis force or any kind of pseudo forces. But if the experiment is done in space in any orbits, we must have to consider some pseudo force and its effect on the phase shift measurements. This is because the laboratory frame w.r.t. us is basically another rotating frame of reference due to the rotation of earth and hence for synchronizing this effect we must have to include pseudo forces due to earth rotation. the diagram can be drawn as follows. \cite{1} 
\begin{figure}[H]
\centering
\begin{minipage}[b]{0.5\textwidth}
    \includegraphics[width=\textwidth]{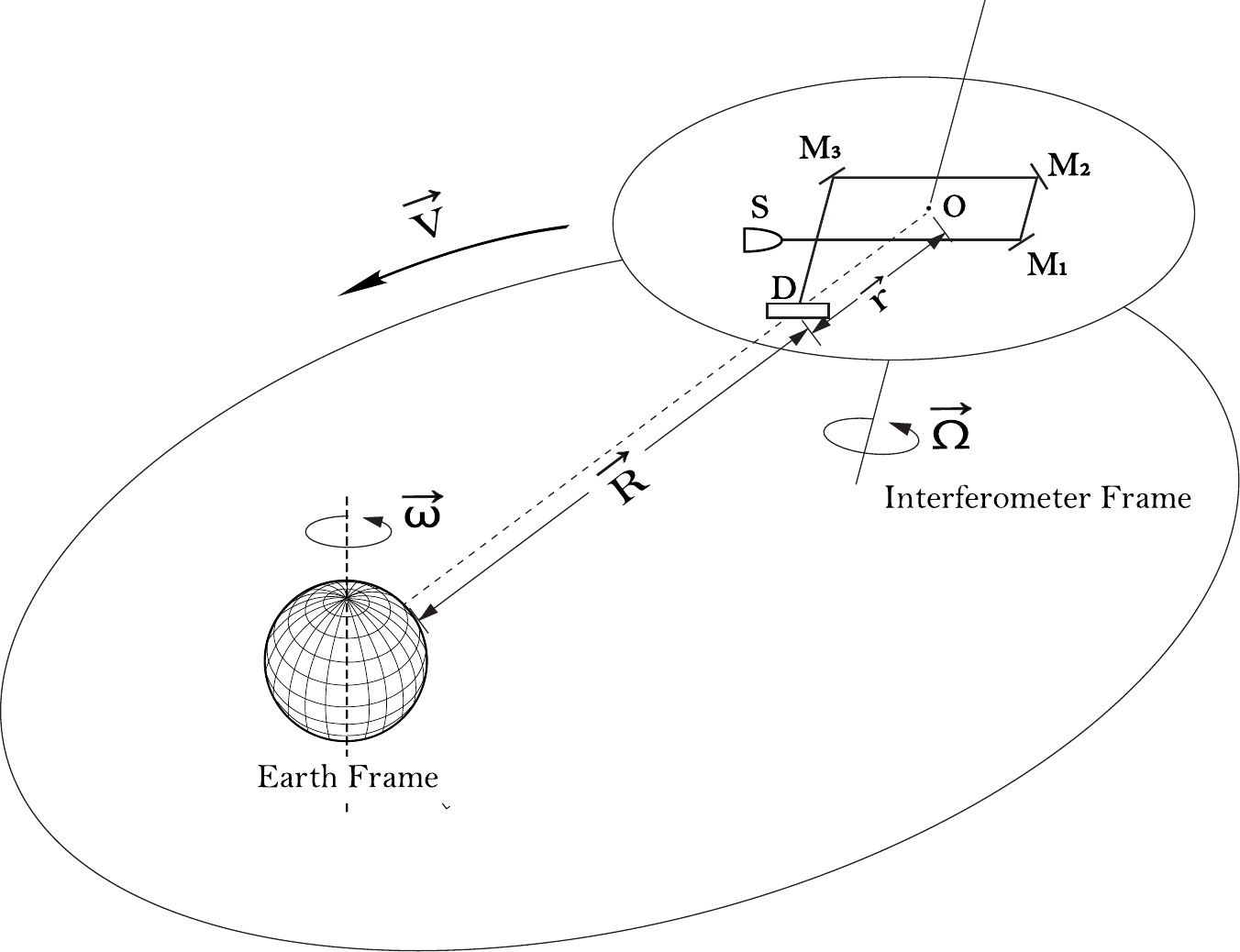}
    \caption{Sagnac Interferometer setup with new geometry in space}
\end{minipage}
\end{figure}
In this orientation of experimental set up we can find total three phase shift. Firstly, the interferometer should have some phase difference between counter propagating waves due to its rotation w.r.t. its center of mass. Second shift should come due to the pseudo effect of the rotating earth frame and its Coriolis effect on rotating gyroscopic interferometer. In this case the phase should oscillate between some positive and negative maximas. The third shift should also come due to pseudo effect of earth rotation on orbital motion of whole interferometric system. This effect should change with the change of radial distance of elliptical orbit. The shift can be discussed mathematically as follows. The total pseudo force can be written as follows in terms of relativistic frame.
\begin{equation}
    F=-\frac{m(\overrightarrow{\omega}\times \overrightarrow{R})+2m(\overrightarrow{\omega}\times \overrightarrow{v})}{1-\frac{v'^2}{c^2}}
\end{equation}
Here $\overrightarrow{\omega}$ is the angular speed of earth frame, $\overrightarrow{v}$ is the velocity of the interferometer cum satellite and $\overrightarrow{R}$ is the radial vector of satellite in rotating earth frame. from this function we can also write the followings,
\begin{equation}
    F_{Coriolis}=-\frac{2m(\overrightarrow{\omega}\times \overrightarrow{v})}{1-\frac{v'^2}{c^2}}
\end{equation}
And,
\begin{equation}
    F_{Centrifugal}=-\frac{m(\overrightarrow{\omega}\times \overrightarrow{R})}{1-\frac{v'^2}{c^2}}
\end{equation}
The term $\overrightarrow{v'}$ is the velocity of satellite on stationary frame and this can be written as follows.
\begin{equation}
    \overrightarrow{v'}=\overrightarrow{v}+(\overrightarrow{\omega}\times\overrightarrow{v})
\end{equation}
In the present configuration of the system, the centrifugal can not provide work done for circular orbit and hence the Coriolis force should provide the phase shift. The potential can be written as follows.
\begin{equation}
    \phi=2m\overrightarrow{R}.(\overrightarrow{v}\times\overrightarrow{\omega})=2mv\omega R\cos{\theta}\sin{\psi}
\end{equation}
Where, $\sin{\psi}=\left |\widehat{\omega}\times\widehat{v}\right |$ and $\cos{\theta}=\left |\widehat{R}.\widehat{n}\right |$. Here, $\widehat{n}=\widehat{\omega}\times\widehat{v}$. For the derivation of phase shift due to this Coriolis effect can be done using Einstein Equivalent Principle. According to that principle the perturbed gravitational field should be equivalent to this pseudo force into linearized approximation. Hence, the time delay can be measured with the following relation.
\begin{equation}
    d\tau=\sqrt{(1+\frac{2\phi}{c^2})}dt\approx(1+\frac{\phi}{c^2})dt
\end{equation}
Hence, considering the both counter propagating lights towards beam splitter we get the following formalism. For the light towards motion the time delay should be as follows. (Non-Relativistic Frame)
\begin{equation}
    \tau_1=T+\frac{2mv\omega R\cos{\theta}\sin{\psi}T}{c^2}
\end{equation}
For opposite wave this should be as follows. Here $T$ is the time period for the orbit w.r.t. the stationary frame.
\begin{equation}
    \tau_2=T-\frac{2mv\omega R\cos{\theta}\sin{\psi}T}{c^2}
\end{equation}
Hence the time delay can be measured as follows.
\begin{equation}
    \Delta\tau=\tau_1-\tau_2=\frac{4mv\omega R\cos{\theta}\sin{\psi}T}{c^2}
\end{equation}
In Relativistic frame it should be as follows.
\begin{equation}
    \tau_1=T+\frac{2mv\omega R\cos{\theta}\sin{\psi}T}{c^2(1-\frac{v'^2}{c^2})}
\end{equation}
and,
\begin{equation}
    \tau_2=T-\frac{2mv\omega R\cos{\theta}\sin{\psi}T}{c^2(1-\frac{v'^2}{c^2})}
\end{equation}
Therefore, the resultant time delay can be written as follows.
\begin{equation}
    \Delta\tau=\tau_1-\tau_2=\frac{4mv\omega R\cos{\theta}\sin{\psi}T}{c^2(1-\frac{v'^2}{c^2})}
\end{equation}
here the velocity of satellite w.r.t. the rotating earth frame can be written as a sum of its orbital motion and rotation motion about its axis. Hence this can be formalized as follows.
\begin{equation}
    \overrightarrow{v}=\overrightarrow{v}_1+(\overrightarrow{\Omega}\times\overrightarrow{R})
\end{equation}
Here, $\overrightarrow{v}_1$ is the orbital velocity w.r.t. rotating earth frame and $\overrightarrow{\Omega}$ is the angular speed of the satellite about its own axis w.r.t. to rotating earth frame. The term $\overrightarrow{\Omega}\times\overrightarrow{R}$ can provide oscillating velocity of the system in rotating frame of reference. Hence the final version of time delay can be written as follows in both Non-relativistic and Relativistic frame a follows.
\begin{equation}
    \Delta\tau=\tau_1-\tau_2=\frac{4mT\overrightarrow{R}.((\overrightarrow{v}_1+(\overrightarrow{\Omega}\times\overrightarrow{R}))\times\overrightarrow{\omega})}{c^2}
\end{equation}
For Relativistic frame the velocity in stationary frame can be written as follows.
\begin{equation}
    \overrightarrow{v'}=(\overrightarrow{v}_1+(\overrightarrow{\Omega}\times\overrightarrow{R})+(\overrightarrow{\omega}\times(\overrightarrow{v}_1+(\overrightarrow{\Omega}\times\overrightarrow{R}))
\end{equation}
Hence the time delay in Relativistic frame can be written below.
\begin{equation}
    \Delta\tau=\tau_1-\tau_2=\frac{4mT\overrightarrow{R}.((\overrightarrow{v}_1+(\overrightarrow{\Omega}\times\overrightarrow{R}))\times\overrightarrow{\omega})}{c^2(1-\frac{v'^2}{c^2})}
\end{equation}
Now this phase shift should provide additional spectral shift in polychromatic interferometer in its spectral content interference spectrum. The polychromatic light should feel different Doppler effect due to this Coriolis force. Hence, for the present case the Doppler shift should be as follows.
\begin{equation}
    z=(g_{00})^{-1/2}-1=\frac{1}{\sqrt{(1+\frac{2\phi}{c^2})}}-1
\end{equation}
Now final form of the intensity spectrum can be formalized as follows for both in relativistic and non relativistic frame using the phase relation $\Delta\phi_{phase}=\frac{2\pi}{\lambda}\Delta\tau$ and satellite radius $r$.
\begin{equation}
\begin{split}
    I_0 (\lambda)=H_0 \exp{(-\frac{1}{2}(\frac{\lambda-\lambda_0}{\Delta\lambda})^2)(1+\frac{1}{\sqrt{(1+\frac{2\phi}{c^2})}})^2}G'(\lambda)\\(1+\cos{(\frac{4mT\overrightarrow{R}.((\overrightarrow{v}_1+(\overrightarrow{\Omega}\times\overrightarrow{R}))\times\overrightarrow{\omega})}{c^2(1-\frac{v'^2}{c^2})})})
\end{split}
\end{equation}
And for non relativistic case we must have the following relation.
\begin{equation}
\begin{split}
    I_0 (\lambda)=H_0 \exp{(-\frac{1}{2}(\frac{\lambda-\lambda_0}{\Delta\lambda})^2)(1+\frac{1}{\sqrt{(1+\frac{2\phi}{c^2})}})^2}G'(\lambda)\\(1+\cos{(\frac{4mT\overrightarrow{R}.((\overrightarrow{v}_1+(\overrightarrow{\Omega}\times\overrightarrow{R}))\times\overrightarrow{\omega})}{c^2})})
\end{split}
\end{equation}
The asymmetric and excess term $G'(\lambda)$ can be written as follows.
\begin{equation}
\begin{split}
    G'(\lambda)=1\\+ \frac{A}{6}((\frac{\lambda-\lambda_0}{\Delta\lambda})^2(1+\frac{1}{\sqrt{(1+\frac{2\phi}{c^2})}})^2\\-3(\frac{\lambda-\lambda_0}{\Delta\lambda})(1+\frac{1}{\sqrt{(1+\frac{2\phi}{c^2})}}))\\-\frac{C}{24}((\frac{\lambda-\lambda_0}{\Delta\lambda})^4(1+\frac{1}{\sqrt{(1+\frac{2\phi}{c^2})}})^4\\-6(\frac{\lambda-\lambda_0}{\Delta\lambda})^2(1+\frac{1}{\sqrt{(1+\frac{2\phi}{c^2})}})^2+3)
    \end{split}
\end{equation}
Where the conditions for $G(\lambda)$ can also be written like this.
\begin{center}
    $G'(\lambda) \rightarrow
    \begin{cases}
      =1 & \text{for symmetric wave form spectrum} (A=B=0)\\
      \neq 1 & \text{for asymmetric wave form spectrum} (A\neq B\neq 0)
    \end{cases}$  
\end{center}

\section{Concluding Remarks}
The Sagnac effect on relativistic frame with the effect of pseudo force have been discussed here in this paper. The primary structure of the interferometer is a square type Sagnac interferometer. The relativistic phase shift have been introduced in the interference spectrum. Relativistic Doppler shift has been incorporated in the spectrum analysis. Equation 20 represents the input Doppler shifted broadband intensity profile. From equation 21 we observed that the asymmetry and excess of the non isotropic gaussian spectrum has been shifted due to relativistic motion of the gyroscopic frame. Finally equation 22 represents the interference spectral content output. Equation 2 represents same intensity output with proper time frame. Here, this effect or phase shift has also been discussed as a result of pseudo effect and Einstein equivalence principle.\par
The general theory of relativity has provided the idea of time delay due to linearized gravity and weak gravity approximations. This weak approximation can be used to measure the pseudo effect on any orbital satellite due to rotating earth frame. Hence, this idea has been incorporated into our new synchronization methods. The figure 3 explains a satellite based Sagnac interferometer. There are three motion in that figure. Firstly, the phase shift is happening due to the axial rotation of the satellite in space. Secondly, that phase shift is modifying due to Coriolis effect of the rotating earth frame. These Coriolis effects have come on the orbital motion and angular motion of the satellite about rotating earth frame. the Coriolis effect has been incorporated to measure the phase shift through time delay using linearized gravity theories. Equation 39 is the time delay for the pseudo force discussed here. For our set up centripetal force do not providing any additional effect. Equation 42 represents the interference spectrum output with this effect. Finally, the Sagnac effect has been established for pseudo effect in terms of Einstein Equivalence Principle.

\section*{Limitations and Future scope of this work}
We have established the theory on Sagnac Effect on Relativistic Frame (Both General Relativity and Special Relativity) and its equivalence with Einstein gravity. The simulated and experimental results analysis will be done in future.

\section*{Acknowledgement}
Shouvik Sadhukhan thanks to Prof. C S Narayanamurthy for important discussion on this work and Acknowledges the SERB/DST(Govt. Of India) for providing financial support via the  project grant CRG/2020/003338 to carry out this work.

}

\end{multicols}

\end{document}